# Quantifying the impact of clinical-academic collaborations

Mohamad Zeina[1], Nick McNally[2], Karl S. Peggs[2,3], and Parashkev Nachev[1]

[1] High Dimensional Neurology Group, UCL Queen Square Institute of Neurology, University College London, Queen Square, London, UK

[2]Research Directorate, NIHR University College London Hospitals Biomedical Research Centre, London NW1 2PG, UK.

[3]Research Department of Haematology, UCL Cancer Institute, 72 Huntley Street, London, WC1E 6DD, United Kingdom

## Abstract

**Background.** Academic collaboration is of self-evident value. But a quantitative representation of its characteristics is needed if policy decisions on its optimal form are to be adequately supported. No established methodological approach to such representation exists. Here we introduce the application of graphical and bibliometric analysis of open data to the task of quantifying the impact of academic networks, with NIHR Biomedical Research Centre (BRC) clinical-academic partnerships in England as the prototype, establishing a general framework for the objective characterisation of academic collaborations and their downstream impact.

**Methods.** We define publication-level bibliometric identities for the 20 English BRCs based on the conjunction of authors from each BRC's partner institutions, yielding a robust network-level unit of analysis for each collaboration. Drawing on OpenAlex and Altmetric records from 2007 to 2025 and the NIHR's current register of infrastructure-supported projects, we characterise the graphical properties of each network, estimate what the university adds to the hospital's papers, what the partnership adds to the papers of relatively infrastructure-poor collaborating institutions, and how that gain depends on existing infrastructure. We use the NIHR UCLH/UCL BRC as an example of the application of the framework to contextualising the performance of one network across others in the same domain.

**Results.** There is substantial variation in the global properties of the 20 networks. UCLH/UCL authored 20,985 network papers from April 2007, in collaboration with 9,868 distinct external partners over the whole record, forming the most central node of the graph of networks across England. Within its hospital, university co-authored papers exhibited 1.6 times the field-weighted citation impact (FWCI) of hospital-only papers, 68% more citations after adjustment for year, field and team size, and were 2.1 and 2.3 times as likely to be cited by a patent or reported in the news. Across 40 UK healthcare partner organisations in sites lacking NIHR early-phase infrastructure, joint publications with the exemplar reached 3.5 times the FWCI of the same organisations' papers written without it (bootstrap 95% CI 2.4 to 4.9; median ratio 3.6), and were 1.8 times (1.5 to 2.0) as likely to reach the top decile of their field as the same organisations' papers written with other networks. Across 60 of the exemplar's most partnered with UK healthcare organisations, the benefit rose from 1.8 times where local NIHR infrastructure activity was densest to 3.4 times where it was sparsest, while impact without the exemplar varied little (mean FWCI 2.86, 2.54 and 2.45).

**Conclusions.** Academic networks can be robustly identified from open data, enabling comparative analysis of collaborative impact on publication outcomes. Applied to England's NIHR BRCs, the approach enables quantification of the impact across networks, and reveals that benefit is most pronounced where infrastructure is least developed. Definitions, results, and code are released for replication and extension.

## Introduction

Translational science is a collaborative enterprise, perhaps no more so than in the domain of healthcare where clinical-academic partnerships arguably constitute the fundamental prototype of research organisation [1]. This is reflected in UK government policy, the National Institute for Health and Care Research (NIHR) creating its first Biomedical Research Centres (BRCs) in April 2007, now expanded to a cohort of 20, as local clinical-academic collaborative networks focused on translational research [2,3]. The desired collaborative span has now broadened to include institutions comparatively lacking in infrastructural support [4], reflecting the scope for wider distribution of UK research [5].

The welcome development of collaboration as a distinct infrastructural attribute demands principled, ideally quantitative, characterisation of its forms. While clinical outcomes are better at research-active hospitals [6,7], and collaborative papers exceed the individual kind in citation rates [8–10], the impact of neither the fact, nor the specific pairing, of clinical-academic institutional collaborations has been robustly quantified. There are two principal obstacles to such characterisation. First, academic networks, including the clinical-academic kind, are relatively opaque bibliographically, for the relationship is imperfectly, if at all, captured in publication meta-data. Second, while detailed data on outputs and their impact may be available within-networks, cross-network analysis is inhibited by the limited public availability of comparators faithful across the domain.

Here we overcome these obstacles by attributing a published output to a collaborative network based on the conjunction of the institutions of its authors, defining a bibliographic identity from publication metadata alone that can be assigned to each of the 20 English BRCs. The participating institutions—the host NHS trust sites and their associated university—thereby define an output as non-exclusively belonging to a network where at least one author is present from each side of the clinical-academic divide. This provides a public, objective criterion that can be applied identically across sites, and allows us to relate the attributes of each collaborative network to its published outputs, and to draw comparisons across networks.

Of the array of questions the approach allows us to answer quantitatively we focus on three. First, what is the added value of the academic institution by comparison with the hospital alone? Second, what is the added value to external organisations that collaborate with the network, specifically to healthcare organisations lacking NIHR early-phase infrastructure? Third, does that added value depend on how much infrastructure the collaborating organisation already has, the point on which the benefit of expansion of collaborative infrastructure rests?

Here we apply this framework to all 20 networks, and report the NIHR UCLH/UCL BRC as a fully worked exemplar. The objective is to demonstrate how clinical-academic partnerships can be quantitatively characterised from open data, and to illuminate the magnitude, nature, and variability of the effects of collaborative infrastructure, and their interrelations.

## Methods

### *Data sources*

OpenAlex, an open catalogue of over 270 million scholarly works with author, institution and citation attributes [11], provided publication metadata stored as a local snapshot (OpenAlex snapshot of 30 March 2026, 272.5 million works). Patent, policy-document and news mentions were drawn from Altmetric (Altmetric database extract of February 2025) [12]. The NIHR Open Data dataset *NIHR Infrastructure Supported Projects* was used to identify infrastructure [13]. The primary measure of a paper's scholarly impact was field-weighted citation impact (FWCI), which normalises citation counts against papers of the same field, document type and publication year to enable faithful comparison [14]. FWCI is OpenAlex's own field-normalised indicator, taken from the work record as supplied (the fwci field), in which the reference set is works of the same OpenAlex primary field, document type and publication year; the same OpenAlex primary field is the field classification used as a covariate in the adjusted models.

We used the date of the NIHR's first BRCs, 1 April 2007, as the first time-point of possible impact. Analyses of output and citation accrual use papers published from 1 April 2007 onward. Analyses of reach and network structure use the whole OpenAlex record for each network because partnerships formed before 2007 remain part of the network being described. Analyses of what partner organisations gain use 2015 to 2024, the period for which Altmetric coverage is stable. The window of every table and figure is stated in its caption. The 2007 window is applied on the publication date, from 1 April 2007, not on the publication year.

### *Establishing BRC bibliographic identity*

In common with other repositories, OpenAlex records author institutions but not their membership in collaborative networks. For each of the 20 English NIHR BRCs in the 2022 to 2027 funding round we therefore constructed a bibliographic identity composed of the host NHS trust paired with the lead university, each identified by its OpenAlex institution identifier (Supplementary Table S1). A paper was non-exclusively attributed to a network when it had at least one author affiliated with the hospital trust and at least one author affiliated with the corresponding university. Each trust was expanded to its individual sites, defined as the trust together with every institution OpenAlex records as its child. The university side is the single university identifier. Five further pairs (King's College London with King's College Hospital, Edinburgh with NHS Lothian, and three US pairs) that are not BRC networks are also derived for comparison.

Reviewing OpenAlex's institution matching we noted nearly every UCLH paper is also identified as a UCL paper, perhaps because the strings share the first three words. We therefore used the raw affiliation strings as written by the authors to classify every distinct string for every pair with a greedy longest-match span classifier, counting an author as a university author only when the written affiliation genuinely names the university rather than one of its hospitals. The classifier was checked against 1,000 affiliation strings independently by a large language model into four resolution mechanisms, with agreement assessed on the binary university-versus-hospital call, and a small subset adjudicated by hand. Second, universities with multiple BRCs were treated as distinct networks: Great Ormond Street Hospital (GOSH) and Moorfields Eye Hospital are treated as separate networks from UCLH/UCL, and Oxford Health from Oxford, and their papers are excluded from the host network even when they would otherwise satisfy its rule. A paper can still count for more than one network when it has authors from both hospitals alongside the shared university, because each network is assessed independently.

We have selected the 2022 to 2027 BRC cohort, evaluating the entire available temporal span. Two centres differ from the previous round: Guy's and St Thomas' with King's College London held a BRC from 2007 to 2022 and is not in the current cohort, and Exeter was not a BRC until December 2022. We use the current cohort throughout and note the consequences in the Discussion.

### *Network structure*

For each of the networks thereby defined, we counted every unique paper and every distinct external institution appearing on those papers, using OpenAlex institution identifiers. To characterise the relationship across the networks themselves, we built an England-wide graph in which each network is a single node (22 nodes: the 20 BRC networks plus King's and Edinburgh) and an edge between two networks is weighted by the number of papers that feature in each. Eigenvector centrality was computed on the weighted graph [15] and betweenness centrality on the unweighted graph [16], using NetworkX [17]. We also computed, for each network, the share of its external partner set (excluding the 40 core university and hospital identifiers) that the exemplar—the NIHR UCLH/UCL BRC—also reaches.

### *The value of clinical-academic partnership*

To quantify the impact of the university's involvement in a hospital's research, we divided each network's partner papers into two groups: hospital-only, lacking a corresponding university-affiliated author, and with-university, with at least one such author. Outcomes were defined as log(1 + citations), FWCI, and binary indicators of at least one patent citation, at least one policy-document citation, and at least one news mention. Each outcome was modelled by ordinary least squares on the group indicator with fixed effects for publication year and primary field and a term in log(1 + author count), with HC3 heteroscedasticity-robust standard errors. The binary outcomes were fitted as linear probability models, so their coefficients are percentage-point differences. The same model was fitted for all 20 networks. Unadjusted ratios of means and of incidence are reported in addition.

### *Comparing sites served vs unserved by NIHR early-phase infrastructure*

We used the NIHR Infrastructure Supported Projects dataset to identify 114 named NIHR infrastructure centres. Biomedical Research Centres, Clinical Research Facilities, HealthTech (formerly MedTech) Research Centres, Experimental Cancer Medicine Centres and BioResource centres were classed as early-phase translational infrastructure; NIHR's applied and delivery schemes (Applied Research Collaborations, Health Protection Research Units and patient-safety centres) were not. The last two early-phase schemes are absent from the dataset, so their centres were taken from the official lists on their own websites (NIHR HealthTech Research Centres, https://www.nihr.ac.uk/about-us/what-we-do/infrastructure/healthtech-research-centres; ECMC network, https://www.ecmcnetwork.org.uk; both accessed 12 September 2026). Every city hosting at least one early-phase centre was treated as served, and organisations based anywhere else were treated as in unserved places. The served set comprises 27 cities: the 25 identified from the dataset and the official lists, plus Guildford and Preston, whose Clinical Research Facilities were found in the NIHR open data on a later pass (6 August 2026). Every result in this paper uses the 27-city set. Because NIHR funds England only, all organisations in Scotland, Wales and Northern Ireland are treated as unserved sites; an England-only count is reported as a sensitivity, together with a conservative served set, a BRC-city-only served set, and a variant capping consortium papers with more than 50 institutions.

For each network we counted its distinct UK partner institutions in unserved sites and the subset that are NHS or healthcare organisations. Among the latter we defined an established partnership as one with five or more joint papers and a founding paper, counted those active (at least one joint paper) in 2021 to 2024, measured distances from the network, and, for the cohort of partnerships whose first joint paper fell in 2007 to 2015, the share still active in 2021 to 2024.

### *Collaborative benefit to underserved organisations*

To quantify the benefit to an unserved-site organisation, we compared the impact of its papers co-authored with the network against its own papers written without it, on papers published in 2015 to 2024. The primary sample included every UK healthcare organisation in an unserved site with at least 50 co-authorships with any of 20 English university-hospital pairs (the 19 non-UCL networks plus King's College London with King's College Hospital), so that membership does not depend on collaboration with the UCLH/UCL BRC. This yielded 40 organisations and 86 983 papers. Papers were assigned to three groups: with the exemplar, with another of the 20 pairs but not the exemplar, and with neither. The pooled ratio of mean FWCI with the exemplar to mean FWCI without was computed on work-deduplicated papers with a 2,000-replicate organisation-clustered bootstrap for its 95% interval; medians and per-organisation ratios (for organisations with at least 10, 15 or 20 papers in each arm) are reported alongside. To distinguish the effect of this particular hub from that of elite collaboration in general, we computed the rate at which papers in each group reach the top decile of FWCI in the sample, and the rate ratios of the exemplar arm against the other-pair arm and the neither arm, with organisation-clustered bootstrap intervals.

A second, earlier sample of 14 organisations selected by at least 50 co-authorships with the exemplar itself was used for the four-endpoint analysis (FWCI, raw citations, patent citation, policy citation). For each endpoint we report the unadjusted with-versus-without ratio and an adjusted estimate obtained by residualising the outcome on publication year, primary field, author count and an organisation code with five-fold cross-fitted gradient-boosted regression, then regressing the outcome on the with-exemplar indicator and the cross-fitted prediction with organisation-clustered standard errors. This adjustment is a flexible covariate correction, not a fixed-effects or doubly robust causal estimator. The same analysis was repeated for 24 non-BRC NHS trusts located in served cities as a comparison group.

### *Dependence of benefit on local infrastructure*

Finally, we addressed the question of whether the benefit depends on the partner organisation's local research infrastructure. For the 60 UK healthcare organisations that co-publish substantially with the exemplar, a set that spans 18 BRC host trusts, 27 other trusts in served cities and 15 organisations in unserved places, we computed four axes: the volume of NIHR-supported project activity near the organisation (the NIHR activity score: each centre's count of supported studies in the 2022/23 to 2024/25 register, discounted exponentially with distance so that a centre 25 km away counts about a third as much as one on site and a centre 100 km away about 2%, and summed over all early-phase centres), the distance to the nearest BRC, regional gross value added per head, and a catchment-weighted index of multiple deprivation (the last two for the 41 English organisations only). Organisations were divided into tertiles on each axis, and within each tertile the with-versus-without ratio of mean FWCI was computed with organisation-clustered bootstrap intervals on the two axes that showed a gradient. We estimated the interaction between the with-exemplar indicator and the standardised axis in the adjusted model, and computed the rank correlation between organisation-level ratio and scarcity.

### *Statistical reporting and reproducibility*

All analyses were applied identically to all 20 networks, and descriptive metrics for every network are given in Supplementary Table S2 and the within-hospital contrast for every network in Supplementary Table S3. Confidence intervals are 95% throughout, from organisation-clustered bootstrap for ratios and from HC3 or cluster-robust standard errors for regression coefficients. Analyses were performed in Python with pandas, statsmodels [18], scikit-learn [19] and NetworkX [17]. The network definitions with site families, every derived table behind the figures, and the scripts that produce them are available at https://github.com/MohamadZeina/collaboration_UCL; each table carries a manifest entry naming the producing script, its window and its checksum.

## Results

### *Global comparative analysis*

Considerable variation was observed across the 20 networks in the number of outputs and citations (Figure 1; Supplementary Table S2), and their evolution over time (Figure 2). Examining the England-wide graph across networks revealed UCLH/UCL as the node with the highest weighted eigenvector centrality—a measure of a node's influence on the graph weighted by the importance of its neighbours—of the 20 (0.599), with GOSH second (0.569) and Oxford third (0.237).

A

Distinct external partner institutions

UCL/UCLH 9,868
Oxford 9,386
Cambridge 7,103
GOSH 7,028
Imperial 7,028
Manchester 5,913
Nottingham 5,279
Newcastle 5,196
Barts 4,981
Leeds 4,919
Southampton 4,845
Leicester 4,657
Birmingham 4,383
Bristol 4,338
Sheffield 3,747
Maudsley 3,736
Royal Marsden 3,461
Moorfields 3,303
Oxford Health 2,165
Exeter 1,508

0 2000 4000 6000 8000 10000
Institutions

B

Eigenvector centrality, contracted pair network

UCL/UCLH 0.599
GOSH 0.569
Oxford 0.237
Moorfields 0.201
Cambridge 0.182
Imperial 0.171
Southampton 0.121
Newcastle 0.121
Barts 0.105
Manchester 0.102
Maudsley 0.093
Birmingham 0.091
Leicester 0.085
Leeds 0.084
Nottingham 0.072
Sheffield 0.072
Royal Marsden 0.070
Bristol 0.063
Oxford Health 0.050
Exeter 0.010

0.0 0.1 0.2 0.3 0.4 0.5 0.6
Weighted eigenvector centrality

C

Pair-authored papers

UCL/UCLH 26,480
Oxford 24,616
Cambridge 15,149
Imperial 13,878
GOSH 12,016
Nottingham 11,854
Leeds 11,393
Manchester 10,621
Newcastle 9,741
Southampton 9,327
Leicester 7,742
Bristol 7,382
Birmingham 7,340
Sheffield 7,076
Barts 6,973
Royal Marsden 5,896
Maudsley 5,805
Moorfields 4,677
Oxford Health 2,996
Exeter 705

0 5000 10000 15000 20000 25000 30000
Papers

D

Total citations to pair-authored papers

Oxford 1,537
UCL/UCLH 1,365
Cambridge 927
Imperial 619
Nottingham 457
GOSH 427
Southampton 422
Leeds 374
Manchester 354
Newcastle 347
Leicester 301
Bristol 277
Royal Marsden 275
Sheffield 252
Birmingham 235
Maudsley 220
Barts 215
Oxford Health 185
Moorfields 153
Exeter 26

0 200 400 600 800 1000 1200 1400 1600
Citations (thousands)

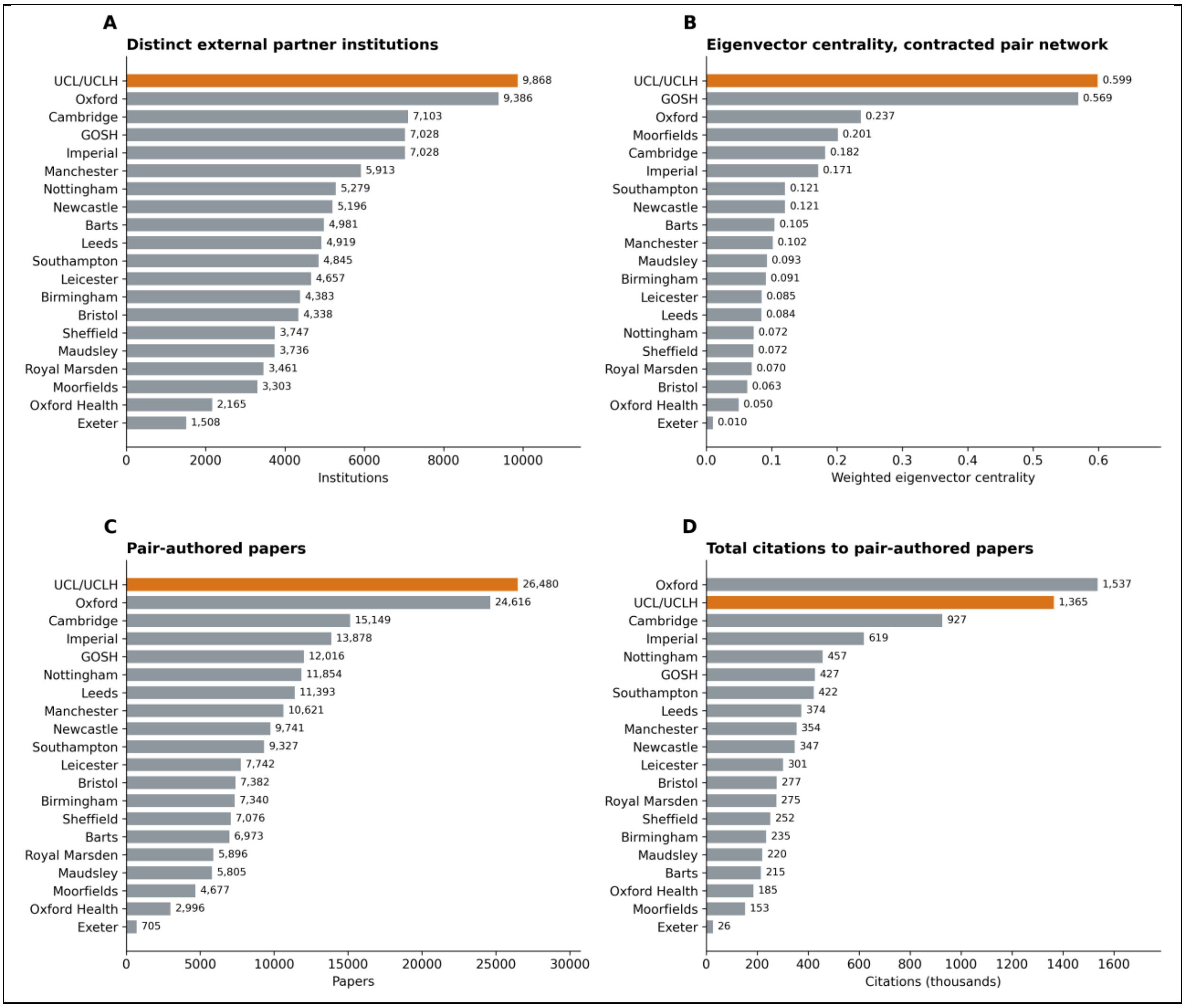

**Figure 1. Global comparative analysis 20 BRC networks (whole OpenAlex record).** (A) Distinct external partner institutions per network. (B) Weighted eigenvector centrality of each network in the contracted pair network. (C) Pair-authored papers per network. (D) Total citations to pair-authored papers. Exemplar (UCLH/UCL) in orange.

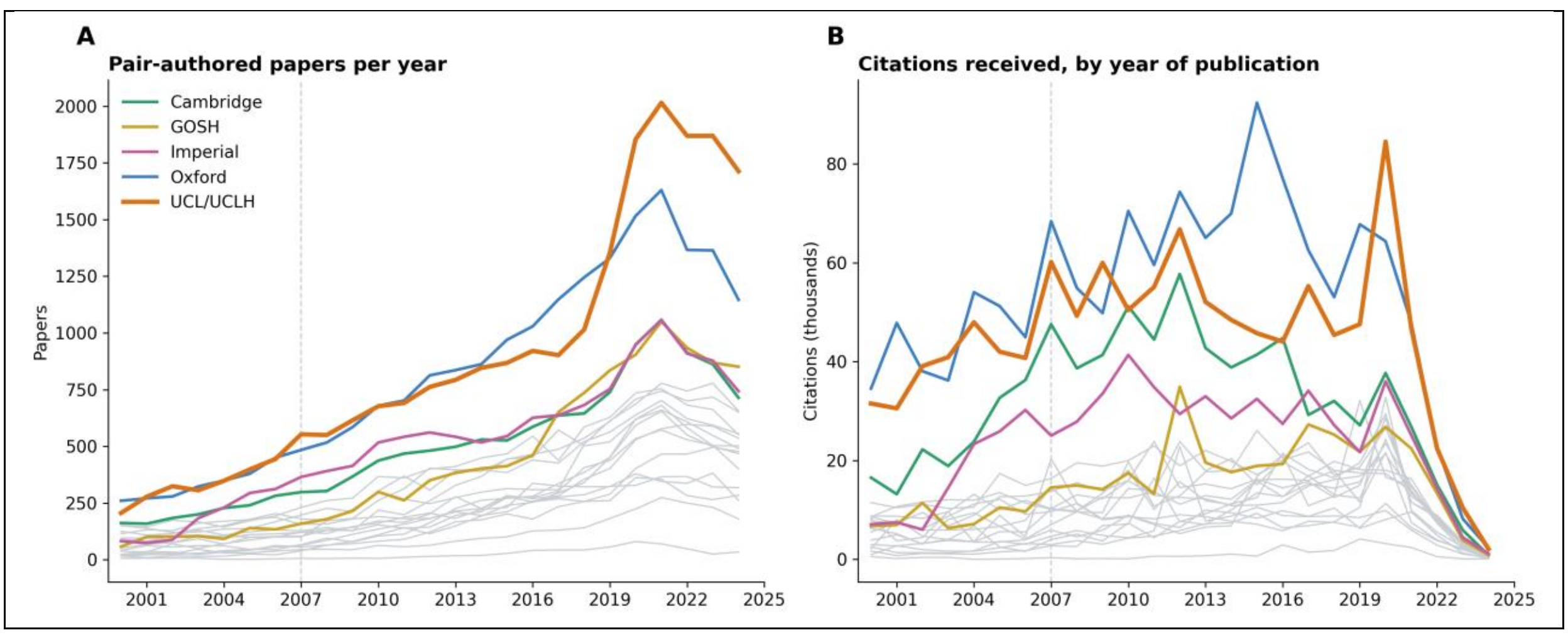

**Figure 2. Output and citation accrual over time.** (A) Pair-authored papers per publication year, 2000 to 2024. (B) Citations received by papers of each publication year. Coloured lines, the five largest networks; grey, the remaining fifteen; dashed line, 2007, the year the first BRCs were established. The fall in recent-year citations reflects accrual time, not a decline in impact.

### *The value of clinical-academic partnership*

Comparison between the exemplar's hospital-only and with-university papers between April 2007 and 2025 showed a substantial difference in volume: 18,504 vs 28,720. More importantly, the mean FWCI of with-university papers was 1.57 times that of hospital-only papers (3.54 against 2.25). Adjusted for publication year, field and team size, they received 68% more citations ($p < 10^{-10}$), 0.85 higher FWCI ($p < 10^{-4}$), a 2.3 percentage-point higher probability of patent inclusion (unadjusted incidence ratio 2.13; $p < 10^{-10}$) and a 6.4 percentage-point higher probability of news coverage (incidence ratio 2.28; $p < 10^{-10}$). With-university papers were, however, less often included in policy documents (incidence ratio 0.78; adjusted difference -0.5 percentage points, $p = 0.01$).

Similar patterns are observed elsewhere (Supplementary Table S3). University co-authorship conferred a significant adjusted citation premium across all 20 networks, ranging from 20% (the Royal Marsden) to 131% (Maudsley), with a median of 58%. A significant adjusted FWCI benefit was seen in 12 of the networks; patent and news effects were significant in 18 and 20 respectively. Policy inclusion rose significantly with university co-authorship in 14 networks and fell significantly in two: the Royal Marsden and UCLH/UCL, both large tertiary centres whose hospital-only output is arguably unusually rich in guideline-facing work.

### *Comparing sites served vs unserved by NIHR early-phase infrastructure*

UCLH/UCL is shown to collaborate with 688 UK institutions in sites without NIHR early-phase infrastructure, more than any other network (Oxford 664, Cambridge 595), among them 418 NHS or healthcare organisations across 237 towns and cities (Oxford 400, Newcastle 395, Cambridge 380; Figure 3A). The lead is maintained under every variant of the definition: England-only (492 institutions and 307 healthcare organisations, against Oxford's 477 and 294), a conservative served set (698 and 424), a BRC-city-only served set (780 and 468), and with consortium papers involving more than 50 institutions removed (625 and 365). It has been first or joint first every year since 2020 (Figure 3B). Since 59% of UCLH/UCL's UK healthcare partners are in unserved places, within 2 points of the highest share of the 20 (Newcastle, 61%) and equal to Oxford's (59%), this effect does not appear to be an artefact of collaboration volume. The largest such partners are University Hospital of Wales in Cardiff (193), Queen Elizabeth University Hospital in Glasgow (158), Ninewells Hospital in Dundee (127), Western General Hospital in Edinburgh (104), Royal Victoria Hospital in Belfast (100) and NHS Greater Glasgow and Clyde in Glasgow (100) (Figure 3D).

The observed pattern is replicated over measures of depth and endurance. Of the exemplar's unserved-site healthcare partnerships, 289 are established, with five or more joint papers (Oxford 260, Cambridge 260), and 286 of those were active in 2021 to 2024 (Oxford 255, Cambridge 254). Of the 226 partnerships whose first joint paper fell in 2007 to 2015, 92% were still producing joint papers in 2021 to 2024, the highest of the 20 (Cambridge 90%, Manchester 89%; median 81%). And they are national rather than metropolitan: 155 established partnerships lie more than 200 km from the exemplar, again the most (Cambridge 134, Imperial 123), at a median distance of 222 km.

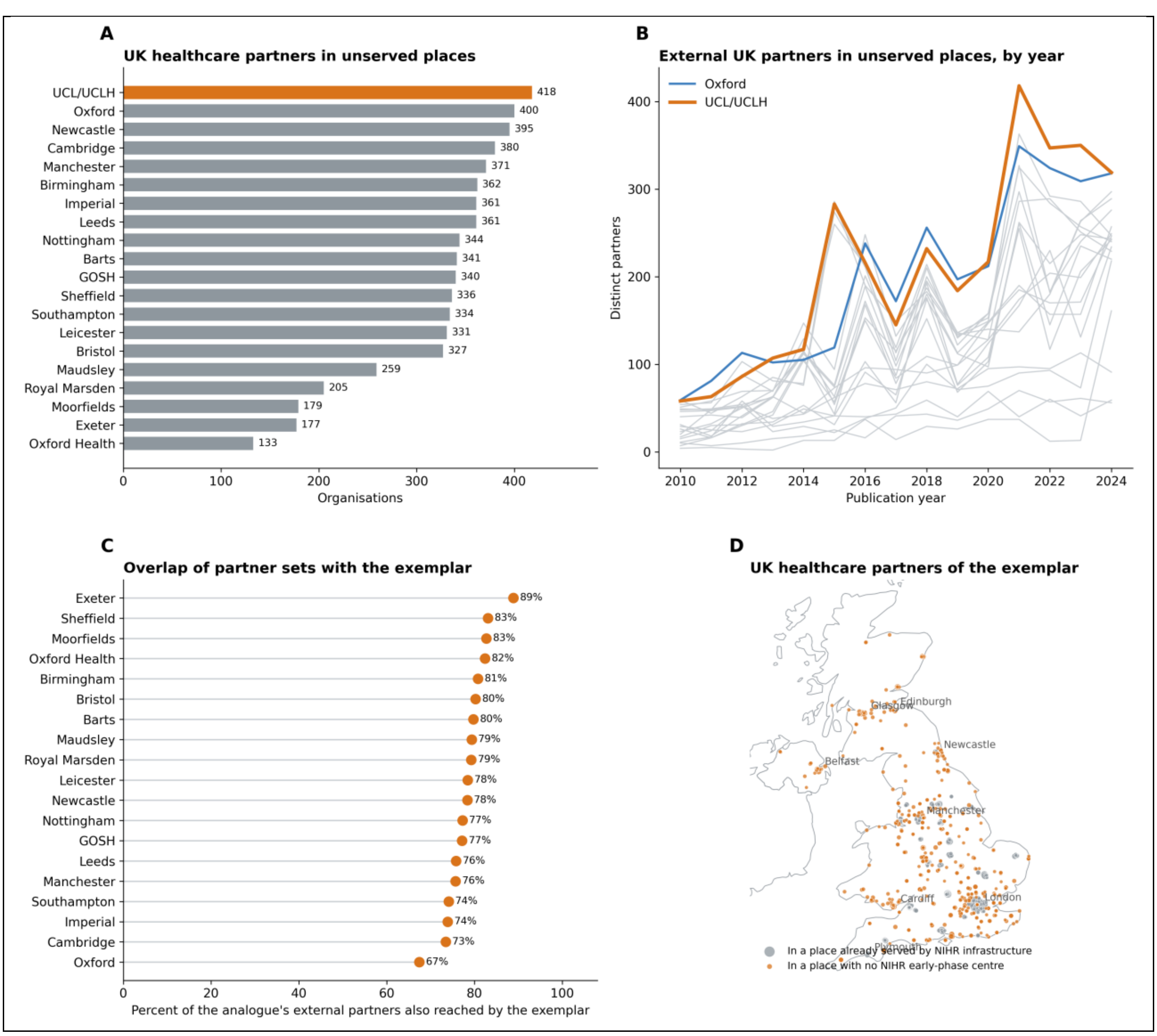

**Figure 3. Reach to healthcare organisations in places without NIHR early-phase infrastructure (whole record; 27-city served set).** (A) Distinct UK healthcare partner organisations in unserved places, per network. (B) Distinct external UK partners in unserved places by publication year; exemplar in orange, Oxford in blue, other networks in grey. (C) Share of each network's external partner set (excluding the 40 core institutions) that the exemplar also reaches. (D) UK healthcare partners of the exemplar, one dot per organisation; orange, in a place with no NIHR early-phase centre; grey, in a served place. Scotland, Wales and Northern Ireland host no NIHR infrastructure because NIHR funds England only.

### *Collaborative benefit to underserved organisations*

Across the 40 unserved-place healthcare organisations in the symmetric sample, papers written with the exemplar exhibited a mean FWCI of 8.02 vs 2.31 without it. The pooled ratio of means was 3.48 (organisation-clustered bootstrap 95% CI 2.36 to 4.93), and 3.57 for the medians (provided since FWCI is heavy-tailed). Among the 36 organisations with at least ten papers in each arm, the median organisation-level ratio was 4.2 and exceeded one in 35 (Figure 4C shows the 14-organisation exemplar-selected sample). The 14 organisations of the original, exemplar-selected sample gave 3.64 (2.28 to 5.46): selecting symmetrically did not weaken the result.

The benefit also exceeded the estimate of what other well-served networks deliver to the same organisations. Of their papers, 30.0% of those coauthored with the exemplar reached the top FWCI decile of the sample, against 16.9% of those coauthored with another of the 20 pairs and 6.7% of those written with neither: rate ratios of 1.78 (1.54 to 2.05) against other pairs and 4.51 (3.95 to 5.18) against neither. For reaching 100 or more citations, the corresponding ratios were 1.66 (1.32 to 2.05) and 4.45 (3.64 to 5.47).

On the four-endpoint analysis of the 14-organisation sample, papers with the exemplar reached 4.04 times the mean FWCI (10.73 against 2.66; adjusted difference 3.69, SE 1.63, $p = 0.02$; 895 against 37,736 papers), 3.19 times the mean citation count (69.7 against 21.8; adjusted log-difference 0.13, $p = 0.17$), 2.59 times the incidence of patent citation (4.3% against 1.7%; adjusted +1.1 percentage points, $p = 0.19$) and 2.60 times the incidence of policy citation (10.0% against 3.8%; adjusted +2.8 percentage points, $p < 10^{-4}$) (Figure 4A). The gain was larger than that of 24 non-BRC trusts in cities that already host early-phase infrastructure (FWCI 3.05, citations 3.11, patents 2.32, policy 2.14; Figure 4B), most clearly on FWCI and policy.

### *The benefit is largest where infrastructure support is scarcest*

Partner organisation papers written without UCLH/UCL delivered similar impact irrespective of the local infrastructure, yielding a mean FWCI 2.86, 2.54 and 2.45 across tertiles of decreasing local NIHR-supported activity. But, strikingly, the with-versus-without ratio rose from 1.83 (1.50 to 2.21) where local NIHR activity was greatest, through 2.68 (1.95 to 4.02), to 3.42 (2.35 to 4.93) where it was lowest (interaction 1.06 per tertile, $p = 0.004$; Spearman $\rho$ between organisation-level ratio and scarcity 0.48, $p < 10^{-4}$; Figure 4D). Regional gross value added per head gave the same gradient (1.87, 2.41 and 3.10 from richest to poorest tertile; interaction $p = 0.007$; $\rho = 0.50$, $p = 0.001$). Distance to the nearest BRC ran the same way without a significant interaction (2.27, 2.05 and 3.50; $p = 0.28$; $\rho = 0.26$, $p = 0.04$), and catchment deprivation showed no gradient ($p = 0.09$). The data suggest that those who gain most from the partnership are those with the least infrastructural support.

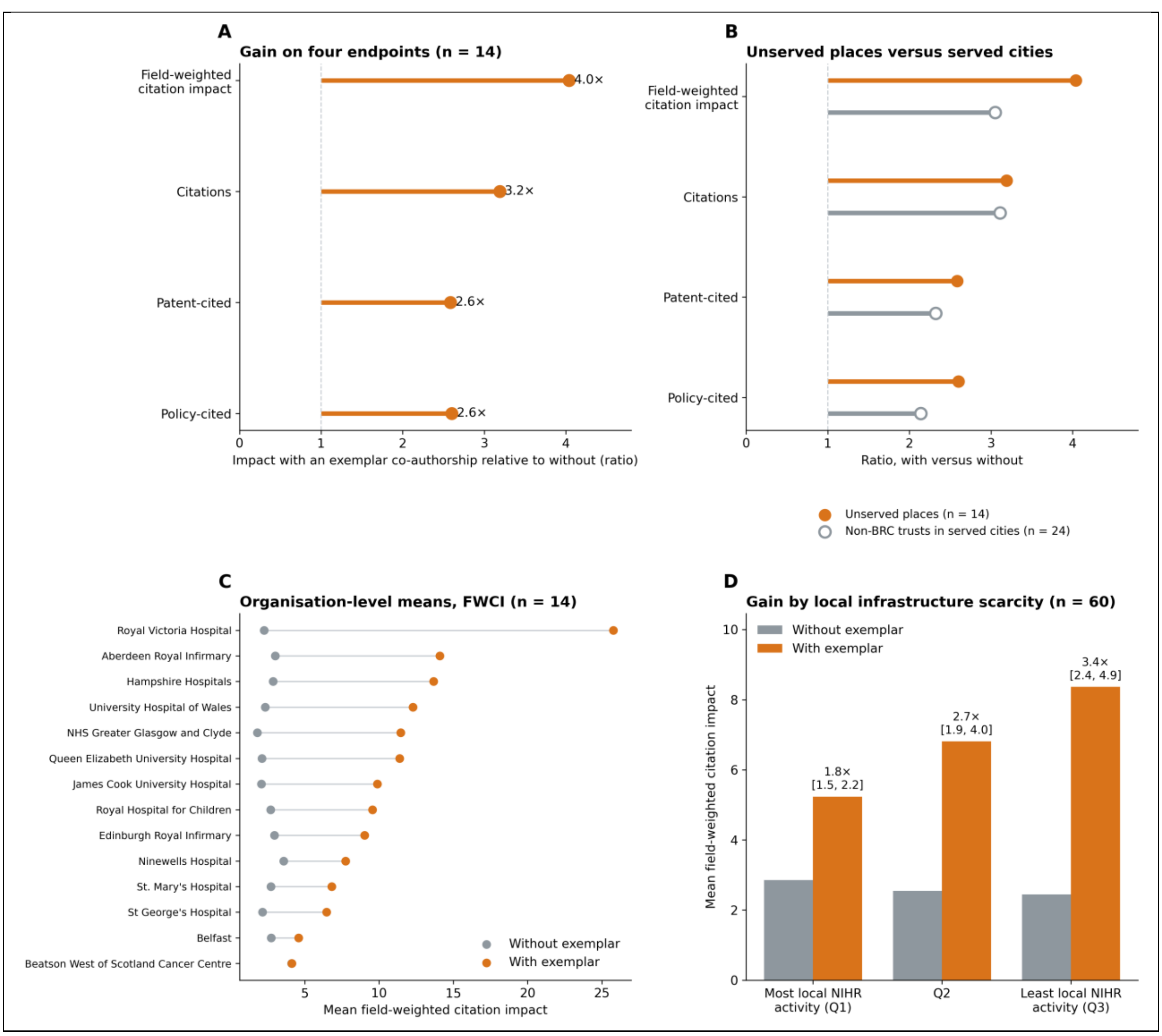

**Figure 4. What unserved-place organisations gain from collaboration with the exemplar (papers of 2015 to 2024).** (A) Ratio of impact of papers with an exemplar co-authorship to papers without, pooled across the 14-organisation sample, on four endpoints; dashed line, no gain. (B) The same ratios for organisations in unserved places (filled, n = 14) and for non-BRC trusts in served cities (open, n = 24). (C) Organisation-level mean FWCI without (grey) and with (orange) an exemplar co-authorship, 14-organisation sample. (D) Mean FWCI without and with the exemplar by tertile of local NIHR-supported research activity, 60 organisations of the exemplar's partner set (20 per tertile); labels give the with-versus-without ratio with organisation-clustered bootstrap 95% interval.

## Discussion

It is beyond doubt that collaboration is essential to many aspects of innovation, and desirable for all. Its optimal form, however, can neither be known *a priori* nor assumed to be the same across disparate domains and contexts. The intelligent formulation of research policy therefore requires a principled means, ideally quantitative, of characterising the attributes of collaborations, so that organisational action can be evidentially motivated. The correct approach to such characterisation is dictated by the essential nature of the activity—a set of relations between entities—naturally formalised as a graph, and effectively described by its properties.

In the setting of inter-institutional collaborations, the first task is identifying the constituent networks. Large-scale comparative analysis necessitates reliance on open data, for reasons of access and comparability. Since research networks are rarely recorded in public scholarly catalogues, network identity must be inferred from publication metadata based on the conjunction of author and institutional identities. Until unique identifiers become ubiquitous this inference will always be imperfect, but adopting a principled, replicable approach exclusively based on open data renders any errors legible and correctable over time. Most importantly, use of the same, transparent criteria of identity ensures different networks can be equitably compared.

In this introductory study, we have employed the simplest criterion of network membership—co-authorship involving at least one representative of the constituent institutions. More complex parameterisation is possible, for example weighting by the number of overlapping authors or their contribution as reflected in author order or correspondence status. The complexity is limited only by the flexibility of the downstream graph modelling framework, to which the scale of available data applies only a relatively weak constraint [20].

Equally, the second task—characterising the network and relating its attributes to outcomes—is open to the full arsenal of available graph-analytic methods. Although here we focus on simple graph properties and employ basic statistical methods, the scale and richness of available data should permit the application of generative models that enable robust meso-scale characterisation, uncertainty estimation, and—most importantly—principled counterfactual analysis, allowing us to move beyond predictive to prescriptive models of collaboration [21].

Applied to the set of clinical-academic networks constituted by the NIHR BRCs in England, our approach quantifies the benefit—over hospital-only research—of involving academic partners across scholarly, enterprise, government, and public attention, revealing substantial variation across the field, plausibly reflective of differences in the ways the networks are constituted, maintained, and exploited. The collaboration benefit does not appear to be merely a function of network size; the precise drivers will require more detailed analysis to determine, with an eye to those most readily amenable to intervention.

Comparison between collaborative partners varying in NIHR support shows that amongst underserved partners the benefit is substantial across all endpoints and, after adjustment, significant for field-weighted impact and policy citation. Strikingly, there is an inverse relation, closely examined in the exemplar, between the magnitude of the benefit and the level of partner NIHR support: UCLH/UCL's most underserved partners gain the most. Were collaboration merely opportunistic, exploiting the strongest work from the strongest peripheral organisations, the collaborative increment would be expected to be either constant or fall with less well-resourced organisations. The data rather suggest that the collaborative contribution here interacts with the partner's needs, engaging more powerfully where the need is greatest.

The way collaborative networks are built and operated may thus be at least as important as their scale and span. Enhancing the infrastructure of an existing effective hub may produce greater benefit to underserved nodes than radically redistributing infrastructural resource; equally, replicating the collaborative policies of highly effective hubs may critically synergize with their planned expansion. As with any complex real-world network, relating structure to outcomes is not straightforward, and demands explicit modelling of a wide array of plausibly material effects that is a fruitful subject of further investigation.

This introductory analysis exhibits an array of limitations. Since co-authorship is inevitably non-random in a way the available data make it difficult to capture, causal effects cannot be inferred, only strong associations. Nonetheless, the within-organisation design accounts for stable differences between organisations, the symmetric sample removes selection on collaboration with the exemplar, and neither the gradient nor the policy reversal is easily explained by selection alone. More sophisticated models, such as of author-level fixed effects comparing the same clinician's papers with and without the hub, or event studies of an organisation's trajectory before and after its first joint paper with a network, may provide further mitigation. It should be noted that the served/unserved classification is purely binary and city-level, approximating effects likely to be more nuanced in reality, and the gradient effects, though statistically significant, are estimated from the exemplar's 60 UK healthcare co-author organisations, most of which hold or sit near early-phase infrastructure of their own. The cohort definition, here based on the 2022-2027 BRCs, is a snapshot of a partially changing field; Guy's and St Thomas' held a BRC for most of the evaluated period but is absent from the roster, while Exeter is present though it joined only in 2022. OpenAlex's institution matching and affiliation coverage are imperfect and have changed over time [22,23]; returning to raw affiliation strings cures the specific artefact we found, but cannot remedy general defects in coverage. Finally, the windows differ between analyses for the reasons given in Methods.

In conclusion, we provide a simple yet effective open-data framework for the quantitative analysis of academic networks that allows faithful comparison between networks and identification of the attributes that predict their impact in bibliometric terms. We demonstrate its application to the clinical-academic partnerships established by the NIHR BRCs, quantifying the benefits of collaboration, and, taking UCLH/UCL as an exemplar, show how the benefit varies with the infrastructural capabilities of constituent partners. Our framework is readily extensible to addressing deeper questions, such as the optimal graphical characteristics of successful networks, with highly expressive models capable of inferring causal effects and providing prescriptive policy guidance.

## Declarations

### *Author contributions*

MZ: conceptualisation, methodology, data curation, analysis, visualisation, writing. NMcN: conceptualisation, funding acquisition, writing. KSP: conceptualisation, funding acquisition, writing. PN: conceptualisation, methodology, supervision, writing.

### *Funding*

This work was supported by the NIHR UCLH Biomedical Research Centre. The views expressed are those of the authors and not necessarily those of the NIHR or the Department of Health and Social Care.

### *Competing interests*

All authors are employed by UCL and/or UCLH and affiliated with the NIHR UCLH BRC, which is one of the 20 centres analysed, and the exemplar reported in full. No other competing interests.

### *Data and code availability*

OpenAlex data are openly available (https://openalex.org). NIHR Infrastructure Supported Projects data are openly available from NIHR Open Data (https://nihr.opendatasoft.com). Altmetric data were used under licence. The network definitions with site families (Supplementary Table S1), every derived table behind the figures and supplementary tables, a manifest recording the producing script, window and checksum of each, and the analysis scripts are available at https://github.com/MohamadZeina/collaboration_UCL. Person-level data are not included.

### *Ethics*

This study used only publicly available bibliographic metadata about institutions and did not involve human participants; ethical approval was not required.

## Supplementary material

Supplementary Table S1. The 20 English NIHR BRC networks (2022 to 2027 funding round). Names as recorded in OpenAlex; some trusts have since merged or been renamed (Bristol's host merged with Weston in 2020; Exeter's host became Royal Devon University Healthcare in 2022). Site families, the OpenAlex child institutions counted as each trust, are given in the repository file network_site_families.csv.

| BRC network | Lead university | Lead hospital trust |
|---|---|---|
| UCLH/UCL | University College London | University College London Hospitals NHS Foundation Trust |
| Oxford | University of Oxford | Oxford University Hospitals NHS Trust |
| Cambridge | University of Cambridge | Cambridge University Hospitals NHS Foundation Trust |
| Imperial | Imperial College London | Imperial College Healthcare NHS Trust |
| Manchester | University of Manchester | Manchester University NHS Foundation Trust |
| Birmingham | University of Birmingham | University Hospitals Birmingham NHS Foundation Trust |
| Bristol | University of Bristol | University Hospitals Bristol NHS Foundation Trust |
| Leeds | University of Leeds | Leeds Teaching Hospitals NHS Trust |
| Barts | Queen Mary University of London | Barts Health NHS Trust |
| Newcastle | Newcastle University | Newcastle upon Tyne Hospitals NHS Foundation Trust |
| Nottingham | University of Nottingham | Nottingham University Hospitals NHS Trust |
| Sheffield | University of Sheffield | Sheffield Teaching Hospitals NHS Foundation Trust |
| Southampton | University of Southampton | University Hospital Southampton NHS Foundation Trust |
| Leicester | University of Leicester | University Hospitals of Leicester NHS Trust |
| Exeter | University of Exeter | Royal Devon and Exeter Hospital |
| GOSH | University College London | Great Ormond Street Hospital for Children NHS Foundation Trust |
| Moorfields | University College London | Moorfields Eye Hospital NHS Foundation Trust |
| Maudsley | King's College London | South London and Maudsley NHS Foundation Trust |
| Oxford Health | University of Oxford | Oxford Health NHS Foundation Trust |
| Royal Marsden | Institute of Cancer Research | Royal Marsden NHS Foundation Trust |

Supplementary Table S2. Descriptive metrics for the 20 networks. Papers (all years and from April 2007), citations, partners and centrality are on the whole OpenAlex record; unserved-place counts use the 27-city served set; established partnerships have five or more joint papers, active means at least one in 2021 to 2024; retention is the share of the 2007 to 2015 partnership cohort active in 2021 to 2024; overlap is the share of the network's external partners also reached by the exemplar. Sorted by all-years papers.

| Network | Pair papers (all) | Pair papers (2007+) | Citations | Distinct partners | Eigen-vector centrality | Unserved-place partners | Unserved-place healthcare partners | Share of UK healthcare partners unserved | Established | Active 2021–24 | Retention | Overlap with exemplar |
|---|---|---|---|---|---|---|---|---|---|---|---|---|
| UCLH/UCL | 26,480 | 20,985 | 1,365K | 9,868 | 0.599 | 688 | 418 | 59% | 289 | 286 | 92% | — |
| Oxford | 24,616 | 18,967 | 1,537K | 9,386 | 0.237 | 664 | 400 | 59% | 260 | 255 | 86% | 67% |
| Cambridge | 15,149 | 11,495 | 927K | 7,103 | 0.182 | 595 | 380 | 60% | 260 | 254 | 90% | 73% |
| Imperial | 13,878 | 12,168 | 619K | 7,028 | 0.171 | 566 | 361 | 56% | 239 | 234 | 84% | 74% |

| Network | Pair papers (all) | Pair papers (2007+) | Citations | Distinct partners | Eigen-vector centrality | Unserved-place partners | Unserved-place healthcare partners | Share of UK healthcare partners unserved | Established | Active 2021–24 | Retention | Overlap with exemplar |
|---|---|---|---|---|---|---|---|---|---|---|---|---|
| GOSH | 12,016 | 10,484 | 427K | 7,028 | 0.569 | 508 | 340 | 57% | 205 | 202 | 79% | 77% |
| Nottingham | 11,854 | 9,104 | 457K | 5,279 | 0.072 | 543 | 344 | 58% | 161 | 155 | 72% | 77% |
| Leeds | 11,393 | 8,892 | 374K | 4,919 | 0.084 | 557 | 361 | 61% | 219 | 210 | 83% | 76% |
| Manchester | 10,621 | 8,079 | 354K | 5,913 | 0.102 | 557 | 371 | 60% | 228 | 224 | 89% | 76% |
| Newcastle | 9,741 | 8,016 | 347K | 5,196 | 0.121 | 574 | 395 | 61% | 184 | 177 | 73% | 78% |
| Southampton | 9,327 | 7,515 | 422K | 4,845 | 0.121 | 528 | 334 | 58% | 136 | 129 | 71% | 74% |
| Leicester | 7,742 | 6,222 | 301K | 4,657 | 0.085 | 479 | 331 | 59% | 121 | 120 | 78% | 78% |
| Bristol | 7,382 | 5,702 | 277K | 4,338 | 0.063 | 489 | 327 | 58% | 145 | 141 | 87% | 80% |
| Birmingham | 7,340 | 6,087 | 235K | 4,383 | 0.091 | 530 | 362 | 59% | 233 | 224 | 83% | 81% |
| Sheffield | 7,076 | 4,727 | 252K | 3,747 | 0.072 | 491 | 336 | 59% | 190 | 186 | 82% | 83% |
| Barts | 6,973 | 4,651 | 215K | 4,981 | 0.105 | 494 | 341 | 58% | 193 | 192 | 88% | 80% |
| Royal Marsden | 5,896 | 4,723 | 275K | 3,461 | 0.070 | 302 | 205 | 52% | 81 | 80 | 72% | 79% |
| Maudsley | 5,805 | 5,650 | 220K | 3,736 | 0.093 | 387 | 259 | 53% | 47 | 45 | 61% | 79% |
| Moorfields | 4,677 | 4,299 | 153K | 3,303 | 0.201 | 262 | 179 | 49% | 43 | 33 | 24% | 83% |
| Oxford Health | 2,996 | 2,509 | 185K | 2,165 | 0.050 | 229 | 133 | 48% | 21 | 19 | 46% | 82% |
| Exeter | 705 | 572 | 26K | 1,508 | 0.010 | 227 | 177 | 56% | 14 | 13 | 56% | 89% |

Supplementary Table S3. Within-hospital contrast for the 20 networks, papers of April 2007 to 2025, text-based arm definition. Adjusted estimates are from OLS on the with-university indicator with fixed effects for publication year and primary field and log(1 + author count), HC3 standard errors; FWCI difference and policy ratio marked (ns) where $p \geq 0.05$. Incidence ratios for patents, news and policy are unadjusted.

| Network | Hospital-only papers | With-university papers | FWCI ratio | Adjusted citation premium | Adjusted FWCI difference | Patent incidence ratio | News incidence ratio | Policy incidence ratio |
|---|---|---|---|---|---|---|---|---|
| UCLH/UCL | 18,504 | 28,720 | 1.57 | +68% | +0.85 | 2.13 | 2.28 | 0.78 |
| Oxford | 19,184 | 27,363 | 1.61 | +67% | +0.99 | 2.06 | 1.86 | 1.27 |
| Cambridge | 15,828 | 14,894 | 1.70 | +63% | +0.88 | 2.07 | 1.82 | 1.02 (ns) |
| Imperial | 16,078 | 16,183 | 1.31 | +46% | +0.20 (ns) | 1.91 | 1.67 | 1.13 |
| GOSH | 12,667 | 13,239 | 1.48 | +53% | +0.41 (ns) | 1.97 | 2.17 | 1.37 |
| Nottingham | 13,371 | 12,066 | 1.15 | +57% | +0.28 (ns) | 1.55 | 1.48 | 1.11 |
| Leeds | 18,860 | 12,252 | 1.28 | +41% | +0.51 | 1.24 | 1.67 | 1.09 |
| Manchester | 14,522 | 10,096 | 1.30 | +52% | +0.15 (ns) | 1.81 | 1.72 | 1.16 |
| Newcastle | 11,337 | 10,582 | 1.49 | +59% | +0.74 | 1.78 | 1.96 | 1.00 (ns) |
| Southampton | 11,353 | 11,056 | 1.43 | +63% | +0.63 | 1.83 | 1.87 | 1.65 |
| Leicester | 11,083 | 8,333 | 2.05 | +67% | +0.96 | 1.83 | 2.26 | 1.26 |
| Bristol | 9,167 | 7,532 | 1.55 | +69% | +0.83 | 1.42 | 2.03 | 1.55 |
| Birmingham | 10,177 | 7,383 | 1.39 | +50% | +0.42 (ns) | 2.05 | 1.84 | 1.19 (ns) |
| Sheffield | 11,184 | 5,957 | 1.26 | +56% | +0.23 (ns) | 1.58 | 1.58 | 1.32 |
| Barts | 23,762 | 5,963 | 1.68 | +47% | +1.13 | 1.30 | 1.70 | 1.02 |
| Royal Marsden | 10,782 | 6,875 | 1.08 | +20% | -0.58 | 1.64 | 1.41 | 0.74 |
| Maudsley | 2,917 | 6,985 | 2.17 | +131% | +0.59 | 2.29 | 2.72 | 1.34 |
| Moorfields | 4,757 | 5,760 | 1.16 | +27% | -0.36 (ns) | 1.83 | 1.98 | 1.01 (ns) |
| Oxford Health | 2,112 | 3,844 | 1.91 | +86% | +1.26 | 0.98 | 2.46 | 1.53 |
| Exeter | 2,184 | 750 | 1.76 | +83% | +1.56 | 0.62 | 3.21 | 1.14 |